\documentclass[12pt]{article}
\usepackage{amsmath,amsfonts,amssymb,bbm}
\usepackage{tensor}
\usepackage[T1]{fontenc}
\usepackage[utf8]{inputenc}
\usepackage{lmodern}
\usepackage{authblk}
\usepackage[disable]{todonotes}
\usepackage{color}
\usepackage{multicol}
\usepackage{cite}
\usepackage{collref}
\usepackage[margin=3cm]{geometry}
\usepackage{hyperref}

\newcommand{\p}{\partial}

\newcommand{\be}{\begin{equation}}
\newcommand{\ee}{\end{equation}}

\def\nn{\nonumber}

\allowdisplaybreaks[1]
\newcommand\email[1]{\thanks{\href{mailto:#1}{\nolinkurl{#1}}}}

\author[a]{Pujian Mao\email{maopj@ihep.ac.cn}\,}
\author[a,b]{Hao Ouyang\email{ouyangh@ihep.ac.cn}\,}

\affil[a]{\,Institute of High Energy Physics and Theoretical Physics Center for Science Facilities, Chinese Academy of Sciences, 19B Yuquan Road, Beijing 100049, P.~R.~China}
\affil[b]{\,School of Physical Sciences, University of Chinese Academy of Sciences, 19A Yuquan Road Beijing 100049, P.~R.~China}

\title{\bf Note on soft theorems and memories in even dimensions\\}
\date{}

\begin{document}

\maketitle
\thispagestyle{empty}

\begin{abstract}
Recently, it has been shown that the Weinberg's formula for soft graviton production is essentially a Fourier transformation of the formula for gravitational memory which provides an effective way to understand how the classical calculation arises as a limiting case of the quantum result. In this note, we propose a general framework that connects the soft theorems to the radiation fields obtained from classical computation for different theories in even dimensions. We show that the latter is nothing but Fourier transformation of the former. The memory formulas can be derived from radiation fields explicitly.
\end{abstract}

\cleardoublepage

\tableofcontents

\section{Introduction}

Recently, a triangle of exact equivalence relation between three classes of physics has been established \cite{string}  (see, also, \cite{Strominger:2017zoo} for a comprehensive discussion). It was first observed by Strominger and collaborators in gravitational theory \cite{Strominger:2013jfa,He:2014laa,Strominger:2014pwa}\footnote{The connection was already observed in Yang-Mills theory by Strominger \cite{Strominger:2013lka}, but the full triangle relation was first established in gravitational theory.}. The precise ingredients of this triangle relation are Weinberg's soft graviton theorem \cite{Weinberg:1965nx}, gravitational memory \cite{memory,Braginsky:1986ia} and Bondi-Metzner-Sachs (BMS) Supertranslation \cite{Bondi:1962px,Sachs:1962wk,Sachs:1962zza}. Those physical phenomena are connected in the following way: The gravitational memory formula is nothing but a Fourier transformation of the soft graviton factor, which suggests that the classical result is a limiting case of quantum result; the soft graviton theorem is precisely equivalent to the Ward identity of the BMS supertranslation that is supposed to be a spontaneously broken symmetry of the quantum gravity S-matrix; gravitational memory can be understood as a transition between two inequivalent vacua of the gravitational field and the initial and final vacua are related by the specific BMS supertranslation. Shortly, this triangle relation has gained considerable attentions and has been extended to electromagnetism \cite{He:2014cra,Campiglia:2015qka,Kapec:2015ena,Pasterski:2015zua}, Yang-Mills theory \cite{He:2015zea,Mao:2017tey} and other theories \cite{Campiglia:2015kxa,Lysov:2015jrs,Dumitrescu:2015fej,Avery:2015gxa,Campoleoni:2017mbt,Campiglia:2017dpg}.

Originally, the triangle relation was established in four dimension and only Weinberg's pole formula \cite{Weinberg:1965nx} was involved. In reality, soft theorems can be viewed as factorization properties that scattering amplitudes must obey in a low-energy expansion. For gauge and gravitational theory, the universal properties go through the subleading order and subsubleading order respectively in a low-energy expansion in arbitrary dimensions \cite{Bern:2014vva} (see, also, \cite{Low:1954kd,GellMann:1954kc,Low:1958sn,Burnett:1967km,Gross:1968in,Jackiw:1968zza,Bell:1969yw,White:2011yy,Cachazo:2014fwa,Casali:2014xpa,Schwab:2014xua,Kalousios:2014uva,Zlotnikov:2014sva}). With the triangle equivalence in mind, one would expect that any new discovery at one corner will hint at new physics at other corners which are intensively discussed recently. In four dimension, new memory effects were reported \cite{Pasterski:2015tva,Mao:2017axa} inspired by the subleading soft theorems. On the asymptotic symmetry side, the investigation is still ongoing and an agreement has not been reached. The subleading soft theorems can be understood as Ward identities of new symmetries \cite{Kapec:2014opa,Lysov:2014csa,Campiglia:2014yka,Campiglia:2015yka,Campiglia:2016jdj,Campiglia:2016hvg,Campiglia:2016efb,Kapec:2016jld}. But the nature of those new symmetries are not completely known namely it is not clear that the new reported symmetries are asymptotic symmetries of the theories of interest. Alternatively, a consistent picture was proposed \cite{Conde:2016csj,Conde:2016rom} that the subleading soft theorems can be recovered as a Ward identity associated to the same symmetries that control the leading piece of the theorem. The situation in higher dimension is less clear even at the leading order. On the one hand, the asymptotic symmetry (\textit{e.g.} the BMS supertranslation in gravitational theory) that is supposed to be responsible for the leading soft theorem seems to be debatable \cite{Hollands:2003ie,Tanabe:2009va,Tanabe:2011es,Kapec:2014zla,Kapec:2015vwa,Hollands:2016oma}. On the other hand, the memory effect in higher dimension was shown to be absent \cite{Garfinkle:2017fre}.

The aim of the present note is to clarify the connection between soft theorems and memories in even dimensions for different theories. Using the retarded and advanced Green's functions in even dimensions, we obtain the retarded and advanced solutions of the wave functions in Minkowski spacetime. The radiation fields \cite{Dirac:1938nz} are defined by the difference of the retarded and advanced solutions. We find that the radiation fields are nothing but a Fourier transformation of the soft factors of the theory that we are solving for the radiation fields. In the asymptotic region, the memory formulas can be reproduced explicitly by the radiation fields in four dimension. In higher dimensions, we show that in $d$ dimension there is no memory effect at the order $r^{1-d/2}$  in the $1/r$ expansion which is in consistence with the results of \cite{Garfinkle:2017fre}. However there are memories in dimensions higher than four if one is able to trace the subleading orders in the $1/r$ expansion. The nontrivial memory is of order $r^{3-d}$ \footnote{This was also noticed in \cite{Chu:2016qxp,Chu:2016ngc}.}. Consequently, we propose that the equivalence of the soft theorems and memories should be understood in the following way: The classical computation of the radiation fields arises as a limiting case of the quantum results namely soft theorems \footnote{A well known example is from the soft bremsstrahlung process that was discussed in details in section 6.1 of \cite{Peskin:1995ev}.} and the memory effects are determined completely by the radiation fields in the asymptotic region.

The organization of this note is quite simple. In the next section we establish the general set-up. We compute the radiation fields from the classical field equations and show that the the radiation fields are nothing but Fourier transformation of the soft factors. In section 3, we discuss the relation of the radiation fields and the memories. The last section is devoted to discussions on open issues.

\section{The set-up }

The key observation of the equivalence of the soft theorems and memories is to consider the soft factors as the expectation value of fluctuation of fields \cite{Strominger:2014pwa,Pasterski:2015tva,Pasterski:2015zua,Mao:2017axa}. This will be also applied in the present work. We start with the scalar field to establish our general set-up.

\subsection{Scalar field}

Soft scalar theorem has been recently derived in \cite{Campiglia:2017dpg}\footnote{As already pointed in \cite{Campiglia:2017dpg}, the soft scalar theorem is only a tree-level result because scalar fields will not retain massless after including loop corrections.}:
\be
 M_{N+1}^{\mathrm{scalar}}=\sum_{k=1}^N \frac{g_k}{p_k\cdot q} M_{N}^{\mathrm{scalar}}+\mathcal{O}(\omega^0),
\ee
where $g_k$ are the coupling constants, $q$ is the momentum of the soft scalar and $p_k$ are the momenta of the hard particles. We will consider massive hard particles since charged particle sources moving at the speed of light seem to be ill-defined \cite{Tolish:2014bka} when solving the classical wave equations. The soft factor will be interpreted as the expectation value of the scalar field in the process of scattering in momentum space \cite{Strominger:2014pwa,Pasterski:2015tva,Pasterski:2015zua,Mao:2017axa} which leads to\footnote{One may consider $M_{N+1}$ as the expectation value of the scalar fluctuation produced in the process of $n\rightarrow N-n$ scattering while $M_{N}$ may be regarded as the ``vacuum'' expectation value in this process.}
\be
\lim_{\omega\rightarrow0}\omega\,\varphi_{d}(\omega,\vec{q})=\lim_{\omega\rightarrow0} \omega\frac{M_{N+1}^{\mathrm{scalar}}}{M_{N}^{\mathrm{scalar}}},
\ee
where $d$ denotes the dimension of the spacetime. Hence,
\be\label{scalar}
\varphi_{d}(\omega,\vec{q})=\sum_{k=1}^N \frac{g_k}{p_k\cdot q},
\ee
at the low energy limit. Performing a Fourier transformation, one obtains the scalar field in the position space as
\be
\varphi_{d}(x)=\sum_{k=1}^N\int \frac{d^{d-1} q }{(2\pi)^{d-1}}\frac{1}{2\omega}\frac{\eta_k g_k }{q\cdot p_k}(e^{i q\cdot x}+\mathrm{c.c.})
\ee
where $\eta_k=1$ or $-1$ for an outgoing or incoming particle\footnote{We take a different convention than \cite{Campiglia:2017dpg} where all legs outgoing were assumed.}.

We now calculate $\varphi_{d}(x)$ in even dimension $d=4+2n$. It is very useful to define the generating function
\be
\Phi\equiv\sum_{n=0}^\infty \frac{1}{n!}(\pi  s)^n\varphi_{4+2n}.
\ee
Then $\varphi_{d}(x)$ can be derived from the generating function easily by taking the limit
\be
\varphi_{4+2n}(x)=\lim_{s\rightarrow0}\frac{1}{\pi^n}\frac{d^n}{ds^n}\Phi
\ee
The generating function can be solved out explicitly
\be\begin{split}
\Phi& =-\sum_{k=1}^N\sum_{n=0}^\infty\eta_k  g_k  \int_{-\infty}^{\infty} d\omega_q\int_{0}^{\pi}d\theta \int_{0}^{\pi}d\phi \int d\Omega_{2n} \frac{\omega_q^{2n}\sin^{2n+1}\theta\sin^{2n}\phi}{2(2\pi)^{3+2n}n!}(\pi  s)^n
\\&\phantom{=} \times e^{i \omega_q(r\cos\theta -t)}
({E_k-|\mathbf{p}_k| \sin\theta_k\sin\theta\cos\phi- |\mathbf{p}_k| \cos\theta_k\cos\theta })^{-1}
\\
& =-\sum_{k=1}^N\sum_{n=0}^\infty\eta_k  g_k  \int_{-\infty}^{\infty} d\omega_q\int_{0}^{\pi}d\theta \int_{0}^{\pi}d\phi \frac{2n!(4\pi)^n}{(2n)!} \frac{\omega_q^{2n}\sin^{2n+1}\theta\sin^{2n}\phi}{2(2\pi)^{3+2n}n!}(\pi  s)^n{}
\\&\phantom{=} \times e^{i \omega_q(r\cos\theta -t)}
({E_k-|\mathbf{p}_k| \sin\theta_k\sin\theta\cos\phi- |\mathbf{p}_k| \cos\theta_k\cos\theta })^{-1}
\\&=-\sum_{k=1}^N\int_{-\infty}^{\infty} d\omega_q\int_{0}^{\pi}d\theta \int_{0}^{\pi}d\phi \frac{\eta_k  g_k  }{(2\pi)^{3}}
\frac{\cos(\omega_q \sqrt{-s}\sin\theta\sin\phi)\sin\theta e^{i \omega_q(r\cos\theta -t)}}{E_k-|\mathbf{p}_k| \sin\theta_k\sin\theta\cos\phi- |\mathbf{p}_k| \cos\theta_k\cos\theta }\\
&=-\sum_{k=1}^N\frac{\eta_k g_k }{2}\int_{-\infty}^{\infty} d\omega_q\int_{0}^{\pi}d\theta \int_{-\pi}^{\pi}d\phi \frac{\sin\theta}{(2\pi)^{3}} \exp(i \omega_q(r\cos\theta - t + \sqrt{-s}\sin\theta \sin\phi))\\
&~~~~\times(E_k-|\mathbf{p}_k| \sin\theta_k\sin\theta\cos\phi-|\mathbf{p}_k| \cos\theta_k\cos\theta )^{-1}.
\end{split}\ee
Here we consider $\Phi$ as a function of $(t,r,\theta_k)$.
The signature of the metric is  $(-,+,...,+)$.
It is useful to note that one can write this generating function as a three dimensional integral over the subspace spanned by $\mathbf{x}$, $\mathbf{p_k}$ and $\mathbf{n}$ with $\mathbf{n}\cdot \mathbf{x}=\mathbf{n}\cdot \mathbf{p}=0$ and $|\mathbf{n}|=1$:
\be
\Phi=\sum_{k=1}^N\int \frac{d^{3} q }{(2\pi)^{3}}\frac{1}{2\omega}\frac{\eta_k g_k }{q\cdot p_k }(e^{i q\cdot (x+\sqrt{-s}n)}+\mathrm{c.c.}),
\ee
where $n^\mu=(0,\mathbf{n})$. It follows that
\be
\Phi=\sum_{k=1}^N g_k  \Phi_k \label{softscalar},
\ee
where \footnote{ The field (\ref{generating}) has the spacetime origin as a special point. From the point of view of the soft theorems, spacetime translation corresponds to a phase shift $\exp(i q \Delta x)$ in the soft factor.}
\be\label{generating}
\Phi_k= \frac{\eta_k}{4\pi\sqrt{(x\cdot p_k)^2-p_k^2 x^2+p_k^2 s}} (\Theta(t-\sqrt{r^2-s})-\Theta(t+\sqrt{r^2-s})).
\ee
As we will see shortly that this is nothing but the radiation field obtained from the solutions of the massless scalar wave equation.

In $d=4+2n$ dimensional spacetime, the retarded and advanced Green's functions satisfying the equation
\be\label{wave}
-\eta^{\mu\nu}\p_\mu\p_\nu G_{d}(x-x')=(\p_t^2-\sum_{i=1}^{3+2n}\p_{x_i}^2) G_{d}(x-x')=\delta_{d}(x-x'),
\ee
are given by \cite{Hassani}
\begin{align}
G^{\mathrm{ret}}_{4+2n}&=\frac{1}{2 \pi^{n+1}}\delta^{(n)}((t-t')^2-|\mathbf{x}-\mathbf{x'}|^2)\Theta(t-t'),\label{regreen}\\
G^{\mathrm{adv}}_{4+2n}&=\frac{1}{2 \pi^{n+1}}\delta^{(n)}((t-t')^2-|\mathbf{x}-\mathbf{x'}|^2)\Theta(t'-t).\label{adgreen}
\end{align}
Considering the source corresponding to a particle created or destroyed at the origin
\be\label{particle}
S_{dk}=\int_0^{\infty}d \tau \delta^d(x^\mu-\eta_k p^\mu_k\tau),
\ee
we obtain the retarded solution for the wave function
\be
-\eta^{\mu\nu}\p_\mu\p_\nu \varphi_{dk}^{\mathrm{ret}}=S_{dk},
\ee
as
\be
\varphi_{4+2n\,k}^{\mathrm{ret}}=\int_0^{\infty} \frac{ d \tau}{2 \pi^{n+1}}\delta^{(n)}(-(x-v_k \tau)^2)\Theta(t-E_k\eta_k \tau).
\ee

Now we introduce the retarded generating function
\be\begin{split}
\Phi^{\mathrm{ret}}_k \equiv & \sum_{n=0}^\infty \frac{1}{n!}\varphi_{4+2n\,k}^{\mathrm{ret}} (\pi  s)^n, \\
=&\int_0^{\infty} \frac{ d \tau}{2 \pi}\delta(-(x-p_k \tau)^2+  s)\Theta(t-E_k\eta_k\tau).\label{sg}
\end{split}\ee
For massive particle source, we have
\be
\Phi^{\mathrm{ret}}_k
=\frac{\Theta(\eta_k (t-\sqrt{r^2- s }))}{4\pi\sqrt{(x\cdot p_k)^2-p_k^2 x^2+p_k^2 s}}.
\ee
The generating function for the advanced solution with massive particle source can be derived in a similar way
\be
\Phi^{\mathrm{adv}}_k
=\frac{\Theta(\eta_k (t+\sqrt{r^2- s }))}{4\pi\sqrt{(x\cdot p_k)^2-p_k^2 x^2+p_k^2 s}}.
\ee
A general source can be written as a linear superposition of such created and destroyed particles \eqref{particle}, so the solutions can be written as a superposition of individual ones. Using the relation $\Theta(\eta_k (t-\sqrt{r^2- s }))- \Theta(\eta_k (t+\sqrt{r^2- s }))=\eta_k[\Theta (t-\sqrt{r^2- s })- \Theta(t+\sqrt{r^2- s })]$ , the radiation field \cite{Dirac:1938nz}
\be\label{classical}
\Phi^{\mathrm{rad}}=\sum_{k=1}^N g_k\left(\Phi^{\mathrm{ret}}_k-\Phi^{\mathrm{adv}}_k\right)
\ee
is precisely the same as \eqref{softscalar}.  {However, \eqref{softscalar} is derived as a low energy limit ($\omega\rightarrow0$) of the quantum computation namely scattering amplitudes while \eqref{classical} is a solution of classical wave equation. The equivalence of those two suggests that the classical calculation arises as a limiting case of the quantum result.}

\subsection{Electromagnetic field}
Now we move on to the fields with spin. Let us first consider the electromagnetic theory. It has been realized long time ago that the scattering of a soft photon displayed universal properties through the subleading order in a low-energy expansion \cite{Bern:2014vva} (see, also, \cite{Low:1954kd,GellMann:1954kc,Low:1958sn} for earlier, restricted to four dimension, versions)
\be
M_{N+1}^{\mathrm{photon}}=\sum_{k=1}^N\left( e_k\,\frac{p_k^\mu }{p_k\cdot q}
+ e_k\,\frac{i J_k^{\mu\nu} q_\nu }{p_k\cdot q}\right)\epsilon_\mu M_{N}^{\mathrm{photon}}+\mathcal{O}(\omega).
\ee
In analogy with the scalar case \eqref{scalar}, one can consider the soft factors to be the classical fields in the momentum space at low energy limit
\be
A^\mu(\omega,\vec{q})=\sum_{k=1}^N\left( e_k\,\frac{p_k^\mu }{p_k\cdot q}
+ e_k\,\frac{i J_k^{\mu\nu} q_\nu }{p_k\cdot q}\right).
\ee
Evaluating its Fourier transformation
\be
A_d^\mu(x)=\sum_{k=1}^N\int \frac{d^{d-1} q }{(2\pi)^{d-1}}\frac{\eta_k}{2\omega}\left( e_k\,\frac{p_k^\mu }{p_k\cdot q}
+ e_k\,\frac{i J_k^{\mu\nu} q_\nu }{p_k\cdot q}\right) (e^{i q\cdot x}+\mathrm{c.c.}),
\ee
it is straightforward to get
\be\label{radphoton}
A_d^\mu(x)=\sum_{k=1}^N e_k(p_k^\mu+J_k^{\mu\nu}\p_{\nu}) \varphi_{dk},
\ee
where $\varphi_{dk}$ is defined by
\be
\varphi_{dk}(x)=\lim_{s\rightarrow0}\frac{1}{\pi^n}\frac{d^n}{ds^n}\Phi_k.
\ee
We will show that \eqref{radphoton} is just the radiation fields obtained from Maxwell's equations in Lorentz gauge:
\be\label{maxwell}
\p^\nu\p_\nu A_d^\mu=- j_d^\mu,
\ee
where $j_d^\mu$ is a conserved current.

A generic current for a collection of charged free point particles associated with the soft factor up to subleading order is given by
\be\label{current}
j^\mu_{d}(y)=\sum_{k=1}^N\,e_k\left[\int_0^{\infty}d\tau p_k^\mu\delta_d(y-\eta_k p_k \tau) + \int_0^{\infty}d\tau  J_k^{\mu\nu}\p_\nu\delta_d(y-\eta_k p_k \tau)\right],
\ee
where the second term of the right hand side is the dipole contribution. Conservation of the current implies conservation of the charge
\be
\sum_{k=1}^N\,\eta_k e_k=0.
\ee

In Lorentz gauge, the Maxwell's equations \eqref{maxwell} take the form of a wave function \eqref{wave} of each component of $A^\mu$. Hence, we can immediately write down the retarded and advanced solutions with the help of the Green's functions \eqref{regreen} and \eqref{adgreen}. Finally, we get
\be
A^{\mu\,\mathrm{ret}}_d=\sum_{k=1}^N e_k(p_k^\mu+J_k^{\mu\nu}\p_{\nu}) \varphi^{\mathrm{ret}}_{dk},\quad A^{\mu\,\mathrm{adv}}_d=\sum_{k=1}^N e_k(p_k^\mu+J_k^{\mu\nu}\p_{\nu}) \varphi^{\mathrm{adv}}_{dk}.
\ee
Then, the radiation field of electromagnetic theory
\be
A^{\mu\,\mathrm{rad}}_d=A^{\mu\,\mathrm{ret}}_d-A^{\mu\,\mathrm{adv}}_d,
\ee
recovers the soft factor in position space \eqref{radphoton} explicitly.

\subsection{Linearized gravitational field}
In this section, we demonstrate our last example which is the linearized gravitational theory. The soft graviton theorem goes through the subsubleading order \cite{Bern:2014vva,Schwab:2014xua,Kalousios:2014uva,Zlotnikov:2014sva} (see, also, \cite{Gross:1968in,Jackiw:1968zza,White:2011yy,Cachazo:2014fwa} for earlier, restricted to four dimension, versions)\footnote{we are using natural units where $8\pi G_N=1$.}
\be
M_{N+1}^{\mathrm{graviton}}=\sum_{k=1}^N \left(  \frac{p_k^\mu p_k^\nu}{p_k\cdot q}
 +\frac{i p_k^\mu J_k^{\nu\alpha} q_\alpha }{p_k\cdot q}
 -\frac{1}{2}\frac{J_k^{\mu\alpha} q_\alpha J_k^{\nu\beta} q_\beta}{p_k\cdot q}
 \right)\epsilon_{\mu\nu}M_{N}^{\mathrm{graviton}} +\mathcal{O}(\omega^2).
\ee
We consider the soft factors as the classical field in the momentum space at low energy limit
\be
\bar h^{\mu\nu}(\omega,\vec{q})=\sum_{k=1}^N \left(  \frac{p_k^\mu p_k^\nu}{p_k\cdot q}
 +\frac{i p_k^{(\mu} J_k^{\nu)\alpha} q_\alpha }{p_k\cdot q}
 -\frac{1}{2}\frac{J_k^{\mu\alpha} q_\alpha J_k^{\nu\beta} q_\beta}{p_k\cdot q}
 \right).
\ee
Similar to the scalar and electromagnetic case, the classical field in position space can be obtained simply by a Fourier transformation
\be\label{radgraviton}
\bar h^{\mu\nu}_d(x)=\sum_{k=1}^N (p_k^\mu p_k^\nu + p_k^{(\mu} J_k^{\nu)\alpha}\p_\alpha + J_k^{\mu\alpha}J_k^{\nu\beta}\p_\alpha \p_{\beta})\varphi_{dk},
\ee
where $\Phi_k$ is again defined in \eqref{generating}.

Now we consider the linearized Einstein equations in harmonic gauge
\be
\p_\alpha \p^\alpha \bar{h}_d^{\mu\nu}=- T_d^{\mu\nu},
\ee
where $T_d^{\mu\nu}$ is a symmetric and conserved stress-energy tensor.

A generic stress-energy tensor for a collection of free point particles associated with the soft factor up to subsubleading order is given by
\be\begin{split}\label{s-e}
T^{\mu\nu}_{d}(y)=&\sum_{k=1}^N \big[\int_0^{\infty}d\tau p_k^\mu p_k^\nu \delta_d(y-\eta_k p_k \tau) + \int_0^{\infty}d\tau p_k^{(\mu} J_k^{\nu)\alpha}\p_\alpha\delta_d(y-\eta_k p_k \tau)\\
&~~~+ \int_0^{\infty}d\tau  J_k^{\mu\alpha}J_k^{\nu\beta}\p_\alpha\p_\beta\delta_d(y-\eta_k p_k \tau)\big],
\end{split}\ee
where the last terms of the right hand side are the dipole and quadrupole contributions \cite{Steinhoff:2009tk,Marsat:2014xea}. The linearized Einstein equations also take the form of a wave function \eqref{wave} of each component. Thus, the retarded and advanced solutions are obtained easily as
\be\begin{split}
&\bar h^{\mu\nu\,\mathrm{ret}}_d=\sum_{k=1}^N (p_k^\mu p_k^\nu + p_k^{(\mu} J_k^{\nu)\alpha}\p_\alpha + J_k^{\mu\alpha}J_k^{\nu\beta}\p_\alpha \p_{\beta}) \varphi^{\mathrm{ret}}_{dk},\\
&\bar h^{\mu\nu\,\mathrm{adv}}_d=\sum_{k=1}^N (p_k^\mu p_k^\nu + p_k^{(\mu} J_k^{\nu)\alpha}\p_\alpha + J_k^{\mu\alpha}J_k^{\nu\beta}\p_\alpha \p_{\beta}) \varphi^{\mathrm{adv}}_{dk}.
\end{split}\ee
Finally, the radiation field
\be
\bar h^{\mu\nu\,\mathrm{rad}}_d=\bar h^{\mu\nu\,\mathrm{ret}}_d - \bar h^{\mu\nu\,\mathrm{adv}}_d,
\ee
is nothing but \eqref{radgraviton}.

\section{Radiation field and memory}
We now turn our attention to the memory effect. We will recover the result of \cite{Garfinkle:2017fre} from the radiation field. If one focuses on the future null infinity, the advanced field will not have memory effects on it. Thus our radiation field should be the same as the retarded field. Nevertheless, it is still meaningful to see this directly. Moreover, we will show that there are nontrivial memories in dimensions higher than four from certain subleading order in the $1/r$ expansion.

\subsection{Scalar memory}
The scalar field generated by the source (\ref{particle}) will produce a force on a distant test particle.  The scalar force on a test particle is given by
\be
f^\mu_d =\nabla^\mu \varphi_{d}.
\ee
In the limit of large $r$ with $u=t-r$ finite, we expand scalar field in dimension $d=4+2n$ as
\be
\varphi_{d}=\sum_{m=0}^{\infty} \frac{\varphi^{(m)}_{d}}{r^{1+n+m}},
\ee
where $\varphi_{d}$ can be obtained from the generating function (\ref{generating}). The leading order behavior of $\varphi_{d}$ is $r^{-1-n}$ and the coefficient of leading order is
\be
\varphi^{(0)}_{d}=\sum_k^N\frac{\eta_k g_k}{4\pi(2\pi )^n r^{n+1}\kappa_k}\frac{d^n}{du^n}\theta(u).
\ee
As proved in \cite{Hollands:2016oma,Garfinkle:2017fre}, the leading order does not give momentum memory effect for $d>4$, which agrees with our result because the momentum memory effect is order $r^{-1-2n}$ as we will show later. However, the momentum memory effect is related to the leading order of $\varphi_{d}$ by equation of motion. Using the wave equation of the scalar field, we obtain
\begin{equation}
\p_u \varphi^{(m+1)}_{d}=\frac{n^2+n-m^2-m-\nabla^2_{S^{2+2n}}}{2(m+1)}\varphi^{(m)}_{d},
\end{equation}
where $\nabla^2_{S^{2+2n}}$ is the Laplace-Beltrami operator on $S^{2+2n}$. The radiation field is zero at $u>0$. Then it follows that
\begin{align}
  &\int_{-\infty}^{\infty} du \p_u\varphi^{(n)}_{d}=(\int du)^{n}\prod_{p=0}^{n}\frac{n^2+n-p^2-p-\nabla^2_{S^{2+2n}}}{2(p+1)}\varphi^{(0)}_{d}, \\
   &\int_{-\infty}^{\infty}du \p_u \varphi^{(m)}_{d}=0,~~~0<m<n,\\
   &\int_{-\infty}^{\infty} du\varphi^{(n-1)}_{d}=(\int du)^{n}\prod_{p=0}^{n-1}\frac{n^2+n-p^2-p-\nabla^2_{S^{2+2n}}}{2(p+1)}\varphi^{(0)}_{d},\\
   &\int_{-\infty}^{\infty} du \varphi^{(m)}_{d}  =0,~~~0<m<n-1.
\end{align}
The momentum memory  can be written as
\begin{align}
\Delta P^\mu_{d}&=\int du\nabla^\mu  \varphi_d =-\Delta \varphi_d K ^\mu +  \int du \p_r \varphi_d R ^\mu
+\sum_{i=1}^{2 +2n}\int du \p_{\phi_i} \varphi_d \nabla^\mu \phi_i,\\
\Delta \varphi_d&
=\frac{1}{r^{1+2n}}(\int du)^{n}\prod_{p=0}^{n}\frac{n^2+n-p^2-p-\nabla^2_{S^{2+2n}}}{2(p+1)}\varphi^{(0)}_{d},\\
 \int du \p_r \varphi_d&
=\frac{-2n}{r^{1+2n}}(\int du)^{n}\prod_{p=0}^{n-1}\frac{n^2+n-p^2-p-\nabla^2_{S^{2+2n}}}{2(p+1)}\varphi^{(0)}_{d},\\
 \int du \p_{\phi_i} \varphi_d&
=\frac{1}{r^{2n}}(\int du)^{n}\p_{\phi_i}\prod_{p=0}^{n-1}\frac{n^2+n-p^2-p-\nabla^2_{S^{2+2n}}}{2(p+1)}\varphi^{(0)}_{d},
\end{align}
where
\be
T^\mu=(1,0),~~~R^\mu =(0,\frac{\mathbf{x}}{r}),~~~K^\mu=-\nabla u =T^\mu + R^\mu.
\ee

Therefore the momentum memory of the radiation field in dimension $4+2n$ is of order $r^{-1-2n}$ and is determined by the leading order coefficient $\varphi^{(0)}_{d}$.

 To see this more precisely, we calculate the memory using the generating function. The generating function of the scalar force on a test particle is given by
\be
F^\mu \equiv \sum_{n=0}^\infty \frac{1}{n!}f_{4+2n}^\mu(\pi s)^n =\sum_{k=1}^Ng_k(-\p_u \Phi_k K^\mu +\p_r \Phi_k R^\mu -r^{-1}\p_{\kappa_k} \Phi_k L_k^\mu),
\ee
where
\be
\kappa_k=E_k-\frac{\mathbf{p}_k\cdot\mathbf{x}}{r},~~~L_k^\mu =-r\nabla\kappa_k=(0,\mathbf{p}_k-\frac{\mathbf{p}_k\cdot\mathbf{x}}{r^2}\mathbf{x}).
\ee

We consider the motion of a test particle near null infinity in the time interval $(-u_1,u_1)$ where $u_1\ll r$.
Then the generating function of the scalar memory which is the change in $d$-momentum due to this force is given by
\be
\Delta P^\mu=\int_{-u_1}^{u_1} d u f^\mu= \sum_{k=1}^Ng_k(  P_{ku}  K^\mu +  P_{kr} R^\mu +  P_{k\kappa_k} L_k^\mu).
\ee

  Let us define a new parameter $S=s r^{-2}$ and take the large $r$ limit with $u_1>0$ fixed. Then, the leading term of $\Delta P^\mu$ is of order $r^{-1}$ and independent of $u_1$:
\be
  P_{ku}=- \frac{\eta_k}{4\pi r}\frac{1}{\sqrt{\kappa_k^2+p_k^2 S}}(1+O(\frac{u_1}{r})),\label{Pumemory}
\ee
\begin{multline}
  P_{kr}= \frac{\eta_k }{4\pi r}\frac{S}{\kappa_k ^2-p_k^2-2 E_k  \kappa_k +\left(E_k ^2+p_k^2\right) S}\left(\frac{E_k  \kappa_k +p_k^2}{\sqrt{\kappa_k ^2+p_k^2 S}}+\frac{E_k -\kappa_k }{\sqrt{1-S}}\right)(1+O(\frac{u_1}{r})),\label{Prmemory}
\end{multline}
\begin{multline}
  P_{k\kappa_k}=- \frac{\eta_k }{4\pi r}\frac{1}{\kappa_k ^2-p_k^2-2 E_k  \kappa_k +\left(E_k ^2+p_k^2\right) S}
\left(\frac{E_k  S-\kappa_k }{\sqrt{\kappa_k ^2+p_k^2 S}}+\sqrt{1-S}\right)(1+O(\frac{u_1}{r})).\label{Pkmemory}
\end{multline}
For $d=4,6,8$, the memories are as follows:
\footnote{We would like to thank  Carlo Heissenberg for pointing out typos in (\ref{p8s}), (\ref{p6v}) and (\ref{p8v}) in our previous version.}
\begin{align}
\Delta P_4^\mu=&\sum_{k=1}^N-\eta_k \frac{g_k}{4\pi r\kappa_k} K^\mu,\\
\Delta P_6^\mu=&\sum_{k=1}^N-\eta_k \frac{g_k }{8\pi^2\kappa _k^3 r^3}
(-p_k^2 K^{\mu }+2 \kappa _k^2 R^{\mu }-\kappa _k L_k^{\mu }),\\
\Delta P_8^\mu=&\sum_{k=1}^N-\eta_k \frac{ g_k}{16 \pi^3\kappa_k^5 r^5} \bigg[3(p_k^2)^2 K^{\mu }+ \left(4 E_k \kappa _k^3+4 \kappa _k^4-4 p_k^2\kappa _k^2\right)R^{\mu }\nn\\&- \left(2 E_k \kappa _k^2+\kappa _k^3-3p_k^2 \kappa _k\right)L_k^{\mu }\bigg].\label{p8s}
\end{align}
In four dimension, we obtain a momentum memory effect in the null direction $K^\mu$ as in \cite{Tolish:2014bka}. In higher dimensions, there are momentum memory effects in radial and transverse directions at order $r^{-1-2n}$ of the expansion.

\subsection{Electromagnetic memory}
We now turn our attention to electromagnetic memory. Electromagnetic field associated with the current (\ref{current}) will produce velocity kick on a distant test particle. Expanding in powers of $1/r$ and using the Maxwell's equations, we can write the radiation electromagnetic field as
\begin{align}
  A_{d\mu} &= \sum_{m=0}^{\infty} \frac{A^{(m)}_{d\mu}}{r^{1+n+m}}, \\
  A^{(m)}_{d\mu}&=(\int du)^{m}\prod_{p=0}^{m}\frac{n^2+n-p^2-p-\nabla^2_{S^{2+2n}}}{2(p+1)}A^{(0)}_{d\mu},\\
  A^{(0)}_{d\mu}&=\sum_k^N\frac{\eta_k g_k v_{k\mu}}{4\pi(2\pi )^n r^{n+1}\kappa_k}\frac{d^n}{du^n}\theta(u).
\end{align}
The radiation field agrees with the results of the retarded field derived in \cite{Garfinkle:2017fre} and the electromagnetic memory is determined completely in terms of $A^{(0)}_{d\mu}$. For simplicity, we consider the case that the current \eqref{current} only has the monopole contribution, which will be dominant in the memory \footnote{The dipole contribution will also lead to memory that corresponds to the subleading soft factor \cite{Mao:2017axa}.}.


The generating function for the  field strength is
\be
\mathcal{F}^{\mu\nu}\equiv 2\nabla^{[\mu} \mathcal{A}^{\nu]}=2\sum_{k=1}^N
{e_k}\nabla^{[\mu}\Phi_k p^{\nu]}_k.
\ee
For a test particle with charge $Q$ and $d$-velocity $v^\nu$, the instantaneous kick in $d$-momentum is given by the generating function
\begin{equation}\begin{split}
\Delta P^{\mu}\equiv& Q\int_{-u_1}^{u_1}\mathcal{F}^{\mu\nu}v_\nu,\\
=&Q\sum_{k=1}^N e_k(E_k P_{k r} R^\mu+\kappa_k  P_{k u} R^\mu+E_k P_{k \kappa_k}L_k^\mu- P_{k u} L_k^\mu),
\end{split}\end{equation}
at the rest frame of the test particle. Here $ P_{k u}$, $  P_{k r}$ and $  P_{k \kappa_k}$ are given in (\ref{Pumemory}),
(\ref{Prmemory}) and (\ref{Pkmemory}). Similar to the scalar case, the memory in dimension $4+2n$ is of order $r^{-1-2n}$. The memory in arbitrary even dimension can be obtain by this generating function. Here we give the results in $d=4,6,8$ cases for our only-illustrative purposes. In the limit $s\rightarrow0$, $P_{k r}$ and $  P_{k \kappa_k}$ vanish. Therefore in four dimension, the test particle obtains a change in its $4$-momentum in the transverse direction
\begin{equation}
\Delta P^{\mu}_4 = Q\sum_{k=1}^N \frac{e_k \eta_k}{4 \pi \kappa_k r} L_k^\mu .
\end{equation}
This agrees with the results obtained by Bieri and Garfinkle \cite{Bieri:2013hqa}. In six and eight dimension cases, the electromagnetic force also gives rise to a radial kick memory effect as
\begin{align}
\Delta P^{\mu}_6 &= Q\sum_{k=1}^N\frac{ e_k\eta_k}{8 \pi ^2 \kappa _k^3 r^3 }
 \left[R^\mu\kappa _k(p_k^2-2 \gamma _k \kappa _k)
 +L_k^\mu(\gamma _k \kappa _k-p_k^2)\right],\label{p6v}\\
\Delta P^{\mu}_8 &= Q\sum_{k=1}^N \frac{e_k\eta_k}{16 \pi ^3 \kappa _k^5 r^5 }
 \bigg[R^\mu\kappa _k (-4 \gamma _k \kappa _k^3-4 \gamma _k^2 \kappa _k^2+4 \gamma _k \kappa _k p_k^2-3( p_k^2)^2)\nn\\
 &+L_k^\mu\left(\gamma _k \kappa _k^3+2 \gamma _k^2 \kappa _k^2-3 \gamma _k \kappa _k p_k^2+3 ( p_k^2)^2\right)\bigg]\label{p8v} .
\end{align}

\subsection{Linearized gravitational field and memory}
The gravitational memory effect is a permanent relative displacement of nearby observers induced by gravitational waves. If two nearby observers are initially at rest and separated by spatial displacement $l^\mu$, then gravitational wave changes their separation by
\begin{equation}
\frac{d^2}{d t^2}l^\mu=-R_{ t\nu t}{}^{\mu}l^{\nu}.
\end{equation}

Similar to the scalar and electromagnetic field cases, we can write $\bar h_{d\mu\nu}$ as
\begin{align}
  \bar h_{d\mu\nu} &= \sum_{m=0}^{\infty} \frac{\bar h^{(m)}_{d\mu\nu}}{r^{1+n+m}}, \\
  \bar h^{(m)}_{d\mu\nu}&=(\int du)^{m}\prod_{p=0}^{m}\frac{n^2+n-p^2-p-\nabla^2_{S^{2+2n}}}{2(p+1)}\bar h^{(0)}_{d\mu\nu},\\
  \bar h^{(0)}_{d\mu\nu}&=\sum_k^N\frac{\eta_k g_k v_{k\mu}v_{k\nu}}{4\pi(2\pi )^n r^{n+1}\kappa_k}\frac{d^n}{du^n}\theta(u),
\end{align}
which agrees with the results of \cite{Garfinkle:2017fre} and the gravitational memory of the radiation field is determined explicitly in terms of $\bar h^{(0)}_{d\mu\nu}$. We consider only the monopole contribution in the stress-energy tensor \eqref{s-e}, which will be dominant in the memory \footnote{The dipole contribution also creates memory that corresponds to the subleading soft factor \cite{Pasterski:2015tva}, while the memory associated to the quadrupole contribution of the stress-energy tensor (\textit{i.e} the memory related to subsubleading soft factor) is not known.}.


The gravitational memory is the double integral of nearby observers near null infinity
\be\label{integral2}
\frac{1}{2}\Delta h^{\mathrm{TT}}_{ij}\equiv-\int_{-u_1}^{u_1} du_2 \int_{-u_1}^{u_2} du R_{ titj},
\ee
in the limit of $r\rightarrow\infty$. It is difficult to calculate the generating function for the gravitational
memory. Nonetheless, we can check the memory for $d=4,6,8$ cases. In four dimension, we get
\be
\frac{1}{2}\Delta h^{\mathrm{TT}}_{4ij}=\sum_{k}\frac{\eta_k}{16 r \kappa _k}(R_i R_j p_k^2+2 L_{{ki}} L_{{kj}}-\delta _{{ij}} p_k^2),
\ee
which agrees with  \cite{1987Natur}. The results in six and eight dimensions are given by
\begin{equation}\begin{split}
\frac{1}{2}\Delta h^{\mathrm{TT}}_{6ij}
=&\sum_{k}
\frac{\eta_k}{64 \pi ^{2} r^3 \kappa _k^3}
\bigg[R_i R_j \left(4  E_k ^2 \kappa _k^2+16  E_k  \kappa _k^3+\kappa _k^2 p_k^2-(p_k^2)^2\right)
   +L_{{ki}} L_{{kj}} \left(8  E_k  \kappa _k-4 p_k^2\right)
   \\&
   +(R_j L_{{ki}}+ L_{{kj}}R_i) \left(-12   E_k  \kappa _k^2+3 \kappa _k p_k^2\right)
-\delta_{{ij}} \left(4  E_k ^2 \kappa _k^2+\kappa _k^2 p_k^2-(p_k^2)^2\right)\bigg].
\end{split}\end{equation}
\begin{equation}\begin{split}
\frac{1}{2}\Delta h^{\mathrm{TT}}_{8ij}
=&\sum_{k}
\frac{\eta_k}{192  \pi ^{3} r^5 \kappa _k^5}
\bigg[L_{{ki}} L_{{kj}} \left(36  E_k ^2 \kappa _k^2+12  E_k  \kappa _k^3-36  E_k  \kappa _k
   p_k^2+2 \kappa _k^2 p_k^2+18 (p_k^2)^2\right) \\&
-(R_j L_{{ki}}+L_{{kj}}R_i ) \left(60  E_k ^2 \kappa _k^3+30  E_k  \kappa _k^4-40  E_k  \kappa _k^2 p_k^2+15
   \kappa _k (p_k^2)^2+5 \kappa _k^3 p_k^2\right)  \\&
+R_i R_j(12  E_k ^3 \kappa _k^3{+}138  E_k ^2 \kappa _k^4{-}6  E_k ^2 \kappa _k^2 p_k^2{-}38  E_k  \kappa _k^3 p_k^2{+}9 \kappa _k^2 (p_k^2)^2{+}23 \kappa _k^4 p_k^2+3 (p_k^2)^3) \\&
-\delta _{{ij}} \left(12  E_k ^3 \kappa _k^3+18  E_k ^2 \kappa _k^4-6  E_k ^2 \kappa _k^2 p_k^2+2  E_k  \kappa _k^3 p_k^2-\kappa _k^2 (p_k^2)^2+3 \kappa _k^4 p_k^2+3 (p_k^2)^3\right)\bigg].
\end{split}\end{equation}

\section{Discussions}

We have shown that one can re-interpret the equivalence of the soft theorems and memories as the equivalence between soft theorems and radiation fields where the soft theorems can be considered as a limiting case of quantum results that recovers the classical computations \textit{i.e.} radiation fields. The memory formulas can be reproduced explicitly from the radiation fields near future null infinity. Now we would like to comment on some aspects that arise in previous sections.

It is amusing to find that the subleading orders of soft theorems are related to the multipole contributions of the source of the wave equations. The universal properties of soft emission stop at subleading and subsubleading order in electromagnetic theory and gravitational theory respectively. A very curious point is the radiation fields including multipoles contributions beyond dipole and quadrupole in the current \eqref{current} and the stress-energy tensor \eqref{s-e}. There should be some reasonings that make the monopole and dipole (also quadrupole for gravitational theory) contributions special to be consistent with the soft theorems. We postpone relevant investigation elsewhere.

A puzzling issue to which we do not yet have an answer is that if there is observational effect associated to the memory in lower order in the $1/r$ expansion in higher dimensions. As we have shown in section 2, the formula of the equivalence is, in general, valid in any even dimension while the memory effect is quite sensitive to dimension of the spacetime. However, this may not be surprised. It is well known that the Newtonian limit of general relativity has different effects in different dimensions. For instance, general relativity in $2+1$ dimensions has a Newtonian limit without force between static point masses \cite{Carlip:1998uc}.

A more subtle one is the fact that the radiation fields that we are dealing with are solutions of source-free wave equations. This may be related to the argument that soft particles that we are concerning in the present work are free. There are indeed some other facts that hint such argument. On the one hand, it is well known that the effect of attaching several soft photon or soft graviton to an arbitrary scattering process is just to supply a product of factors of the Weinberg's pole formula, one for each soft particles \cite{Weinberg:1965nx} \footnote{Yang-Mills theory has a very different behavior in multiple soft emissions \cite{Berends:1988zn}.}. Hence, adding multiple soft particles is like adding one by one and soft particles do not see each other namely they are free particles. On the other hand, the soft theorems are connected to asymptotic symmetries in the triangle equivalence. It is reasonable to consider that the soft particles arise from a proper treatment of representation theory of the asymptotic symmetry group \textit{e.g.} BMS particles in gravitational theory. The BMS particles are well studied in three dimension ( see \cite{Oblak:2016eij} for a comprehensive discussion), and gain renewed attentions in four dimension recently \cite{Oblak:2016eij,Oblak:2017ect}. It would be definitely meaningful to have more investigation elsewhere following this direction to explore the nature of soft particles.

\section*{Acknowledgments}

The authors thank Eduardo Conde and Jun-Bao Wu for useful discussions and valuable comments on the draft. The authors would like to thank the
anonymous referee for the suggestions and comments which are very helpful in improving the original manuscript.
This work is supported in part by the National Natural Science Foundation of China (Grant No. 11575202).

\bibliography{ref}

\bibliographystyle{utphys}

\end{document}